\documentclass[aps,pra,twocolumn,showpacs,showkeys]{revtex4-1}
\usepackage{hyperref}
\hypersetup{
    breaklinks=true
    bookmarks=true,         
    unicode=false,          
    pdftoolbar=true,        
    pdfmenubar=true,        
    pdffitwindow=false,     
    pdfstartview={FitH},    
    pdftitle={Orthogonality catastrophe and decoherence in a trapped-Fermion environment},    
    pdfauthor={A. Sindona,J. Goold, N. Lo Gullo,  S. Lorenzo, F. Plastina},     
    pdfsubject={Shake up in ultra-cold atomic traps},   
    pdfcreator={LaTeX},   
    pdfproducer={PP}, 
    pdfkeywords={key1} {key2} {key3}, 
    pdfnewwindow=true,      
    colorlinks=true,        
    linkcolor=blue,          
    urlcolor=blue,        
    filecolor=red,      
    citecolor=blue     
}
\usepackage{dcolumn}
\usepackage{bm}
\usepackage{graphicx}
\usepackage{amsmath}
\usepackage{latexsym}
\usepackage{bbold}
\usepackage{color}
\usepackage{amsfonts}
\usepackage{amssymb}
\usepackage{array}
\usepackage{times}
\usepackage{epsfig}
\usepackage{epstopdf}
\usepackage{setspace}
\usepackage{bbm}
\usepackage{simplewick}
\newcommand{\fourIdx}[5]{
\setbox1=\hbox{\ensuremath{^{#1}}}
\setbox2=\hbox{\ensuremath{_{#2}}}
\setbox5=\hbox{\ensuremath{#5}}
\hspace{\ifnum\wd1>\wd2\wd1\else\wd2\fi}
\ensuremath{\copy5^{\hspace{-\wd1}\hspace{-\wd5}#1\hspace{\wd5}#3}_{\hspace{-\wd2}\hspace{-\wd5}#2\hspace{\wd5}#4}}
}
\newcommand{\thav}[1]{\big\langle{#1}\big\rangle}
\newcommand{\ket}[1]{\left|{#1}\right\rangle}
\newcommand{\bra}[1]{\left\langle{#1}\right|}

\newcommand{\msc}[1]{\text{\textsc{#1}}}
\newcommand{\Exclude}[1]{}
\newcommand{\Include}[1]{#1}
\newcommand{\eF}{\varepsilon_{\msc{f}}}
\newcommand{\rF}{r_{\msc{f}}}

\newcommand{\en}{\varepsilon_{n}}
\newcommand{\er}{\varepsilon_{2r}}
\newcommand{\hw}{\hbar\omega}
\newcommand{\dg}{\dagger}
\newcommand{\p}{\prime}
\newcommand{\rp}{r^{\p}}
\newcommand{\np}{n^{\p}}
\newcommand{\tp}{t^{\p}}
\newcommand{\ts}{t^{\p{\hskip-0.5pt}\p}}
\newcommand{\rs}{r^{\p{\hskip-0.5pt}\p}}
\newcommand{\spind}{(2s+1)}

\newcommand{\HypF}{\fourIdx{ }{2}{ }{1}{F}}
\newcommand{\HypFR}{\fourIdx{ }{2}{ }{1}{\tilde{F}}}

\begin{document}
\date{\today }
\title{Orthogonality catastrophe and decoherence in a trapped-Fermion environment}
\author{A.~Sindona$^{1,2}$, J.~Goold$^{3,4}$, N.~Lo~Gullo$^{4,5}$, S.~Lorenzo$^{1,2}$, F.~Plastina$^{1,2}$}
\affiliation{
$^1$Dipartimento di Fisica, Universit\`a della Calabria, 87036 Arcavacata di Rende (CS), Italy\\
$^2$INFN sezione LNF-Gruppo collegato di Cosenza, Italy\\
$^3$Clarendon Laboratory, University of Oxford, United Kingdom\\
$^4$Physics Department, University College Cork, Cork, Ireland\\
$^5$Quantum Systems Unit, Okinawa Institute of Science and Technology and Graduate University, Okinawa, Japan}
\begin{abstract}
The Fermi edge singularity and the Anderson orthogonality
catastrophe describe the universal physics which occurs when a
Fermi sea is locally quenched by the sudden  switching of a
scattering potential, leading to a brutal disturbance of its
ground state. We demonstrate that the effect can be seen in the
controllable domain of ultracold trapped gases by providing an
analytic description of the out-of-equilibrium response to an
atomic impurity, both at zero and at finite temperature.
Furthermore, we link the transient behavior of the gas to the
decoherence of the impurity, and, in particular to the amount of
non-markovianity of its dynamics.
\end{abstract}
\maketitle
A Fermi gas may be shaken-up by the switching of even a single,
weakly interacting impurity, producing a complete rearrangement of
the many body wave-function that, as a consequence, loses
essentially any overlap with the initial, unperturbed one. This is
the essence of Anderson's orthogonality
catastrophe~\cite{Anderson:67,Anderson&Yuval}, witnessed by the
singular~(edge-like) behavior of the energy distribution of the
impurity induced excitations. An example of how such a many-body
effect comes into play is provided by X-ray photoemission spectra
from most simple metals, where the expected sharp symmetric peak
at the binding energy of a core level is converted into a power
law singularity, as predicted by the
Mahan-Nozi{\`e}res-Dominics~(MND) theory~\cite{Mahan1,Nozieres}.
Similar patterns have been observed in electron emission via X-ray
absorption and Auger neutralization from carbon based
nanomaterials~\cite{AntoEdge}, and for quantum dots~\cite{heyl}.
Fermi edge resonance and orthogonality catastrophe have been also
revealed by non-equilibrium current fluctuations~(shot noise) in
nanoscale conductors~\cite{novotny}, and enter prominently the
physics of phenomena as diverse as the Kondo
effect~\cite{Anderson&Yuval,quattro} and the scattering or
sticking of a low-energy atom or ion on a metal
surface~\cite{scatt,stick}.

Recently, it has been proposed to observe this universal physics
in controllable ultracold atomic setups where the singular
behavior may be probed either in the time domain, by Ramsey
interference type experiments performed on the impurity
atom~\cite{gooldOC}, or in the frequency domain, by
radio-frequency spectroscopy~\cite{knap}. However, an analytic
framework for the case of a \textit{trapped} Fermi gas is lacking.
In this letter we provide such an analytic description, and
discuss the transient response of a harmonically trapped Fermi gas
following the `\textit{sudden}' switching of an embedded two-level
atom excited by a fast pulse. The interaction with the
\textit{local} impurity produces a local quench of the gas, giving
rise to the Anderson catastrophe. We study the Fermi-edge physics
at zero and finite temperature and both in the frequency domain,
by looking at the excitation spectrum of the gas, and in the time
domain, by analyzing the dynamics of the impurity. Thus, we link
the Fermi edge behavior of the excitation energy distribution to
the decoherence of the impurity. In particular, we investigate the
Loschmidt echo~\cite{diciotto,diciassette} and the
non-Markovianity, using recently developed
tools~\cite{breuermeasure,rivas,fisherNM,geo,pinja}, employed so
far to study open systems in different environments, ranging from
spins~\cite{noi} to Bose-Einstein condensates~\cite{haikka}, and
experimentally tested in optical
set-ups~\cite{experimentNM,chiuri}. We find that the
non-Markovianity of the  decoherent dynamics of the impurity
provides a novel interpretation of the essential physics of the
shake-up process.

We consider a gas of non-interacting cold fermions confined by a
one-dimensional trapping harmonic potential of frequency $\omega
$, and described by the Hamiltonian $\hat{H}_{0} = \sum_{n,\xi}
\en \hat{c}_{n\xi}^{\dg} \hat{c}_{n\xi}$, with $\hat{c}_{n\xi}$
being the annihilation operator for the $n$-th single particle
state of energy $\en = \hw (n+1/2)$ and spin $\xi$. We add a
two-level impurity (an atom of a different species from the
trapped component), with internal states $\ket{g}$ and $\ket{e}$
and Hamiltonian $\Hat{H}_{\msc{i}}=\sum_{i=e,g} \epsilon_i \, \ket
i \bra i$, which is trapped in an auxiliary potential and brought
in contact with the Fermi gas. Such a situation can be achieved
using a species selective dipole potential that has a frequency
much greater than the trap which contains the gas, so that the
impurity motion is essentially frozen. We assume that when the
impurity is in the $\ket{g}$ state, it has a negligible scattering
interaction with the gas, hence the Hamiltonian of the composite
system is given by $\hat{H}=\hat{H}_{0} +
\Hat{H}_{\msc{i}}+\hat{V} \otimes \ket{e}\bra{e}$. With the
fermions in their equilibrium configuration, set by $\hat{H}_{0}$,
we suppose the impurity to be quickly excited so that the gas
feels a sudden perturbation $\hat{V}(t)=\hat{V} \theta(t)$ due to
the interaction, assumed to have an $s$-wave like character.

As it is standard in ultracold atoms, at sufficiently low
temperatures, the pseudo-potential approximation for the
interaction is invoked, which amounts to replacing the complicated
atomic interaction potential with an effective short range
potential of strength $V_{0}$,  localized at the minimum of the
harmonic well, which we scale with the trap length $x_{0}$ such
that $V(x)=\pi V_{0}x_{0}\delta (x)$. Due to the parity of the
single particle wave-functions, only the fermions lying in
even-parity states~($n=2r$, with $r=0,1,2,\cdots$) feel the
impurity and are involved in the shake-up process. Explicitly, the
fermion-impurity interaction is given by $\hat{V}=\sum_{r,\rp,\xi}
V_{r\rp} \hat{c}^{\dg}_{2r\xi} \hat{c}_{2\rp\xi}$, where
$V_{r\rp}=V_{0}{\,}(-1)^{r+\rp}{\,}\gamma_r^{1/2}{\,}\gamma_{\rp}^{1/2}$,
and $\gamma_r=2^{-2r} \pi^{1/2} (2r)!/r!^{2}$ (See appendix A). We
label the highest occupied level by $n_F= 2 \rF$, with $\rF$ a
positive integer, so that the Fermi energy reads $\eF = \hw (2 \rF
+1/2)$.

A key quantity for the following is the \textit{vacuum persistence amplitude}
\begin{equation}
\nu _{\beta }(t>0) = \thav{e^{\frac{i}{\hbar}\hat{H}_{0}t}{\,}e^{-\frac{i}{\hbar} (\hat{H}_{0}+\hat{V}) t}}\text{,}
\label{Eq:VacAmp}
\end{equation}
with $\langle{\cdots}\rangle$ denoting the grand canonical average
over the unperturbed fermion state. $\nu _{\beta }(t)$ is the
probability amplitude that the gas will retrieve its equilibrium
state at time $t$, after the switching on of the perturbation and,
as discussed below, its modulus gives the decoherence factor for
the impurity.

The \Exclude{inverse} Fourier transform $\tilde{\nu}_{\beta }(E)$,
subject to the constraint $\nu _{\beta }(t<0)=\nu _{\beta
}^{\ast}(-t)$, gives the excitation spectrum of the gas.
Turning to the interaction picture, we get
\begin{equation}
\nu_{\beta }(t)=\thav{T e^{\frac{1}{i \hbar
}\int_{0}^{t}d\tp\tilde{V}(\tp)}} \text{,} \quad
\tilde{V}(t)=e^{\frac{i}{\hbar
}\hat{H}_{0}t}\hat{V}e^{-\frac{i}{\hbar }\hat{H}_{0}t}\text{,}
\label{IP}
\end{equation}
which, by virtue of the linked cluster theorem, reduces to an exponential sum of connected
Feynman diagrams, $\nu _{\beta }(t)=e^{\Lambda _{\beta}(t)}$, with:\\
\scalebox{0.95}{\includegraphics{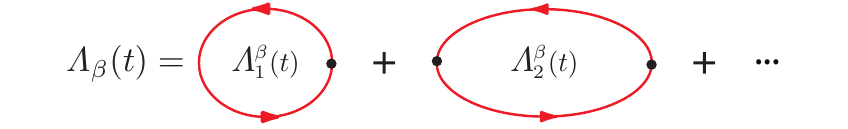}}\\
The closed graphs in $\Lambda _{\beta}(t)$ contain products of
vertices~($V_{r\rp}$) connected by lines~($G_{r}^{\beta }$) that
represent the unperturbed propagators
\begin{equation}
i \hbar G_{r}^{\beta }(t)=e^{-i \er t/\hbar} [ \theta(t)\,f^{-}_{r}
- \theta(-t)\,f^{+}_{r} ], \label{prop}
\end{equation}
where $f^{\pm}_{r}=[1+e^{\pm \beta (\er-\mu)}]^{-1}$ are the
particle-hole distributions, and  $\mu$ denotes the chemical
potential (see Appendix C).

We focus on the lowest order loops, namely
\begin{equation}
\label{Lambda1} \hbar\Lambda _{1}^{\beta }(t)=-i t \, \chi_s
V_{0}\lambda_{+}^{\beta}(0),
\end{equation}
and
\begin{equation}
\label{Lambda2} \hbar^{2}\Lambda _{2}^{\beta }(t)=- \chi_s
V_{0}^{2} \Exclude{\\\notag &\qquad\qquad\qquad\times}
\int_{0}^{t}d\tp\int_{0}^{\tp}d\ts\,\lambda_{+}^{\beta
}(\ts){\,}\lambda_{-}^{\beta }(\ts),
\end{equation}
with $\chi_s=\spind $ accounting for the spin degeneracy and $
\lambda_{\pm }^{\beta
}(t)=\sum_{r=0}^{\infty}{\,}\gamma_r{\,}e^{\pm 2 i r \omega t}
f^{\pm}_{r}$.

This approximation will prove to accurately describe the singular
response of the gas~(contained in the two-vertex term) and to give
the dominant contribution to the shake-up process if the
interaction strength is small in the energy scale of the problem.
The latter is set by both the level separation~$\hw$ and Fermi
energy~$\eF$, which allow us to introduce $\alpha =\frac{\chi_s
V_{0}^{2}}{2\hw\eF}$ as a sensible interaction strength parameter.
\begin{figure}[t]
\centerline{\scalebox{0.98}{\includegraphics{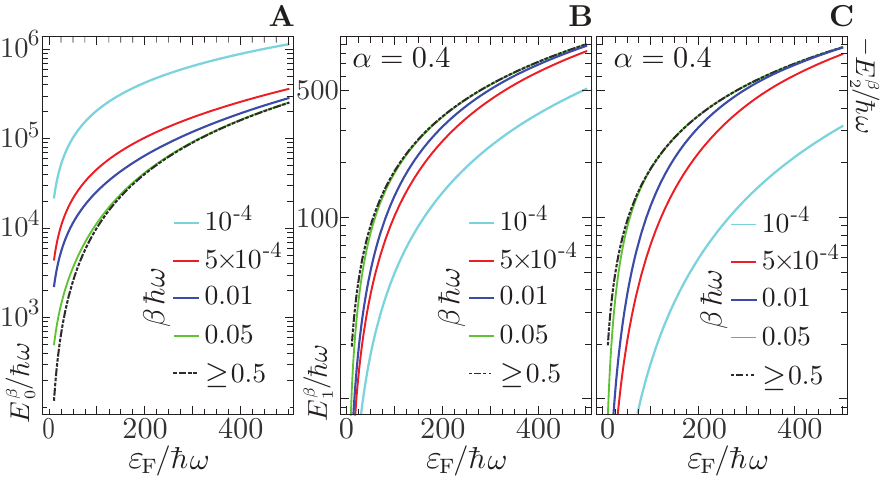}}}
\vskip-12pt \caption{(Color online) Equilibrium
energy~$E^{\beta}_{0}$ of a spin-$1/2$ gas into a harmonic
trap~(panel \textbf{A}) together with the perturbation corrections
$E^{\beta}_{1}$~(Eq.~(\ref{eq:E1}), panel \textbf{B}) and
$E^{\beta}_{2}$~(Eq.~(\ref{Delta2B}), panel \textbf{C}) due to the
effective impurity potential $V(x)$. All Energy curves are
reported in units of $\hw$ \textit{vs} $\eF/\hw$ for different
values of $\beta\hw$ and fixed coupling parameter $\alpha = 0.4$.
\label{eradue}} \vskip-12pt
\end{figure}

\noindent The contribution~(\ref{Lambda1}) may be written as
$\hbar \Lambda _{1}^{\beta }(t)=-i t E_{1}^{\beta} $. Here,
\begin{equation}
E_{1}^{\beta}= \sqrt{2 \chi_s
\hw\eF\alpha}{\,}\sum_{r=0}^{\infty}{\,}\gamma_r{\,}f^{+}_{r}
\label{eq:E1}
\end{equation}
is the first-order shift to the gas energy, as provided by the
Rayleigh-Schr\"{o}dinger perturbation theory. The behavior of the
unperturbed energy $E_{0}^{\beta}=\chi_s
\sum_{n}\en\,f^{+}_{n/2}$, and of its first and second order
corrections (the latter obtained from the two vertex term, see
below) vs $\eF$ is shown in Fig.~\ref{eradue} for various
temperatures. We notice that $E_{0}^{\beta}$ is $1$ to $3$ orders
of magnitude larger than $E_{1}^{\beta }$ for $\alpha \lesssim 1$,
while temperature plays an appreciable role in both
$E_{0}^{\beta}$ and $E_{1}^{\beta}$ for $\beta\hw$ less than $\sim
0.05$.

\begin{figure}[t]
\centerline{\scalebox{0.99}{\includegraphics{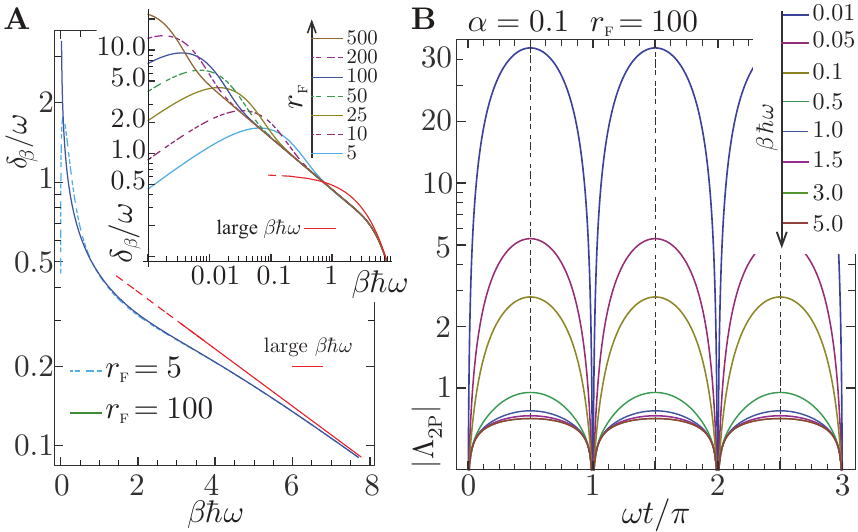}}}
\vskip-10pt \caption{ (color online): (\textbf{A}) Standard
deviation in the Gaussian power law~(\ref{eq:gbeta}), expressed in
$\omega$-units, \textit{vs} $\beta \hw$ for $\rF=5-100$ and
$\alpha=0.1$. The low thermal energy approximation introduced in
the text is also reported. (\textbf{B}) Periodic component
$\Lambda_{2\msc{p}}^{\beta}(t)$ \textit{vs} $\omega t/\pi$, for
$\rF=100$, $\beta \hbar \omega = 0.01-5$ and $\alpha =  0.1$}
\label{eratre}
\end{figure}

While $\Lambda _{1}^{\beta}(t)$ only brings a phase factor to
$\nu_{\beta }(t)$, which corresponds to shifting the spectrum
$\tilde{\nu}_{\beta }(E)$ by $E_{1}^{\beta}$, the two-vertex
connected graph gives the crucial contribution to the persistence
amplitude. As detailed in Appendix B, it can be split into three
parts with well defined trends and physical meaning, i.e.,
$\Lambda _{2}^{\beta}(t)= \Lambda _{2\msc{s}}^{\beta }(t) +
\Lambda_{2\msc{g}}^{\beta }(t) + \Lambda _{2\msc{p}}^{\beta }(t)$.
These represent a (further) energy shift, a gaussian envelope due
to finite temperature effects, and a periodic terms originating
from the equal spacing of the unperturbed single-particle  states,
respectively, and are separately analyzed in
Figs.~\ref{eradue}\textbf{C}, \ref{eratre}\textbf{A}, and
\ref{eratre}\textbf{B}.

The first one, $\hbar\Lambda_{2\msc{s}}^{\beta}(t)=-itE_{2}^{\beta
}$,  provides the second-order correction to the energy of the gas
(the $n>2$-vertex graphs would complete the perturbation series):
\begin{equation}
E_{2}^{\beta } =\alpha \eF \sum\limits_{r{\neq}\rp=0}^{\infty}
\frac{f^{+}_{r}\gamma_r\,\gamma_{\rp}f^{-}_{\rp}}{r-\rp}.
\label{Delta2B}
\end{equation}
Comparing Fig.~\ref{eradue}\textbf{B} and~\textbf{C}, we notice
that the chosen value of $\alpha$ let $E_{2}^{\beta}$ take
absolute values smaller than $E_{1}^{\beta }$. However,
$E_{2}^{\beta}$ is more sensitive to temperature than
$E_{1}^{\beta }$ for $\beta \hw <0.05$.

The second contribution, $
\Lambda_{2\msc{g}}^{\beta}(t)=-\delta_{\beta}^{2} \omega^2
t^{2}/2,\label{Lambda2G}$ produces a Gaussian damping in
$\nu_{\beta }(t)$ and, therefore, a Gaussian broadening in
$\tilde{\nu}_{\beta}(E)$ with standard deviation
\begin{equation}
\delta_{\beta} = \sqrt{2 \alpha g_{\beta }},\quad g_{\beta }=\frac{\eF}{\hw} \sum_{r=0}^{\infty}{\,}\gamma_r^{2}\,f^{+}_{r}\,f^{-}_{r}.
\label{eq:gbeta}
\end{equation}
Here, the coefficient $g_{\beta}$ is weakly influenced by the
Fermi energy, but strongly affected by temperature,  changing by
various orders of magnitude for $\beta \hw \lesssim 0.5$.  No
damping/broadening effects are present at the absolute zero, since
$\delta_{\beta} \to 0$ for $\beta \hw \to
\infty$~(Fig.\ref{eratre}\textbf{A}).

The most important content of the second diagram, giving a non
trivial structure to $\nu_{\beta }(t)$, arises from the third
contribution (see the Appendix):
\begin{equation}
\Lambda _{2\msc{p}}^{\beta }(t)=-\frac{\alpha \eF}{2\hw}\sum\limits_{r{\neq }\rp}^{\infty}\,\gamma_r f_{r}^{+}\,\frac{1-e^{2i(r-\rp)t\omega }}{(r-\rp)^2}\,\gamma_{\rp}f_{\rp}^{-}.
\label{Lambda2E}
\end{equation}
Due to the
harmonic form of trapping potential, this is a periodic function
of time with frequency $2\omega $, see Fig.~\ref{eratre}\textbf{B}.
The zeroes of this sub-graph~(at $\omega t = m \pi$ with
$m=0,\pm1,\pm2$), when combined with the Gaussian
damping~(\ref{eq:gbeta}), yield modulations in the vacuum
persistence amplitude which, as discussed below, are a signature
of non-markovian dynamics of the impurity.

\begin{figure}[h]
\vskip-8pt
\scalebox{0.99}{\includegraphics{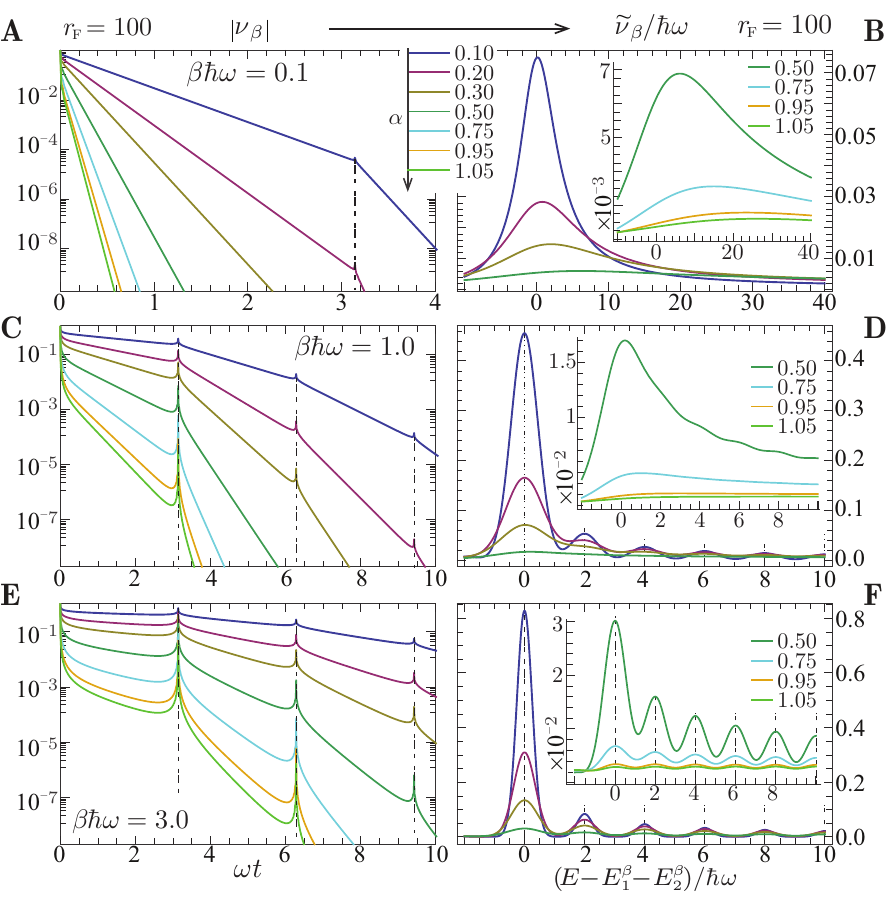}} \vskip-8pt
\caption{(color online): Absolute value of the decoherence factor
$|\nu _{\beta}(t)|$ (left panels,
\textbf{A},\textbf{C},\textbf{E}) and excitation spectrum $\nu
_{\beta}^{\p }(E)$ (right panels,
\textbf{B},\textbf{D},\textbf{F}), calculated from
eq.~(\ref{Vpbet}) by numerically computing the Gaussian
damping~(\ref{eq:gbeta}) and the periodic sub-diagram
$\Lambda_{2\msc{p}}^{\beta}(t)$, for $\beta\hw=0.1-3.0$,
$\rF=100$, and $\alpha =0.1-1.05$.} \label{cinque} \vskip-8pt
\end{figure}

Leaving aside the shifts, the
persistence amplitude is then:\begin{equation}
\nu_{\beta }^{\p}(t)=e^{-\delta_{\beta } \omega^2
t^{2}/2}e^{\Lambda _{2\msc{p}}^{\beta }(t)}\text{.} \label{Vpbet}
\end{equation}
Of particular interest for the discussion below is the behavior of
$|\nu _{\beta}(t)|$ exhibiting spikes at $\omega t \sim \pi ,2\pi
,\cdots $, which become more and more pronounced with increasing
$\beta \hw$, see left panels in Fig.~\ref{cinque}.
The periodicity in the time domain is reflected in the excitation
spectrum $\tilde{\nu}_{\beta }(E)$ that offers an asymmetric,
broadened, signature of the singular behavior of the Fermi gas.
The monotonic structure turns into a sequence of sub-peaks,
separated by $2 \hw$ and related to even-level transitions in the
gas as $\beta \hw$ gets above $\sim 0.5$~(see
Fig.~\ref{cinque}\textbf{B}). These features are observed for any
$\rF$ in the range of $5$ to $100$ (see Appendix C).

The coefficient~(\ref{eq:gbeta}) of the Gaussian power law and the
periodic contribution~$\Lambda _{2\msc{p}}^{\beta }(t)$ can be
approximated as
\begin{equation}
g_{\beta } \approx 2 \sum_{m=1}^{\infty } (-1)^{m} m
\frac{e^{\beta \hw m/2}}{e^{\beta \hw m}-1}, \label{Lambda2GA}
\end{equation}
and
\begin{equation}
\Lambda _{2\msc{p}}^{\beta }(t) \approx \alpha
\sum_{m=-\infty}^{\infty } \ln \frac{e^{2\tau _{m}\omega
}-1}{e^{2(it+\tau _{m})\omega }-1}. \label{LambXX}
\end{equation}
At low temperatures, the leading behavior of the Gaussian standard
deviation is $\delta _{\beta }\approx 2\alpha^{1/2} e^{-\beta
\hw/4}$ for thermal energies $\beta \hw \gtrsim 6$~(see
Fig.~\ref{eratre}\textbf{A}). On the other hand,
Eq.~(\ref{LambXX}) contains a singularity at the absolute zero,
that we regularized by introducing a cut-off parameter $\tau_{0}$.
This regularization is only needed to remove a zero temperature
indefiniteness of the analytic approximation, whereas the
numerical evaluation of the vacuum persistence amplitude does not
suffer from divergence problems. As shown below and as detailed in
the Appendices below, a similar parameter enters the original MND
theory, and we can interpret it as the typical time-scale over
which transitions occur in the gas. On the other hand, thermal
fluctuations introduce other characteristic times
$\tau_{m}=m\beta\hbar$.

\begin{figure}[h]
\scalebox{0.99}{\includegraphics{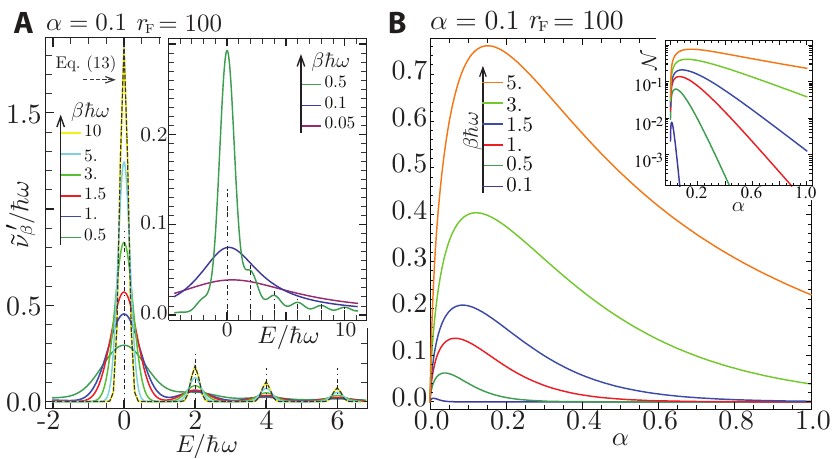}} \vskip -8pt
\caption{(color online): \textbf{(A)}~Absorption spectrum
$\nu^{\p}_{\beta}(E)$, calculated numerically from
eq.~(\ref{Vpbet}) with $\beta\hw=0.1-\infty$, $\rF=100$, and
$\alpha =0.1$, and analytical approximation $\nu_{\beta}^{\p}(E)$
obtained from Eqs.~(\ref{Lambda2GA}) and~(\ref{LambXX}) with $\hw
\beta = 10$ and $\omega\tau_0 = 0.001$;
\textbf{(B)}~Non-Markovianity measure as a function of the
critical parameter $\alpha$ for various temperatures.}
\label{nonmark} \vskip -8pt
\end{figure}

Taking $g_{\beta}$ and $\Lambda _{2\msc{p}}^{\beta }(t)$ as in
Eqs.~(\ref{Lambda2GA}) and (\ref{LambXX}), and using them in Eq.
(\ref{Vpbet}) gives an accurate  approximation to the numerical
results for $\beta \hw \gtrsim 0.1$, number of particles larger
than $10$ and for suitable values of the cut-off parameter, say,
$\omega\tau_0 < 0.02$ (see Fig.~\ref{nonmark}\textbf{A}). In
particular, at $T=0$, the vacuum persistence amplitude takes the
form:
\begin{equation}
\nu_{\beta \to \infty }^{\p}(t) \approx
\left[\frac{e^{2\tau_{0}\omega }-1}{e^{2\omega
(\tau_{0}+it)}-1}\right]^{\alpha} \text{.} \label{Vbet0}
\end{equation}

To compare our findings to the one-dimensional free-fermion
theory, one needs to fix $\alpha $ and let the harmonic frequency
go to zero by keeping the number of particles in the gas~($2\rF
\approx \eF/\hw$) finite. No Gaussian damping occurs in this case,
and the two vertex graph tends to
\begin{equation}
\Lambda_{\msc{mnd}}(t)=-\alpha \ln (it/\tau _{0}+1),
\label{MNDLamb}
\end{equation}
yielding the Nozieres-De Dominicis propagator
$\nu_{\msc{mnd}}(t)=\left( it/\tau _{0}+1\right)^{-\alpha}$,
originally calculated for a suddenly switched on core-hole in a
free electron gas~\cite{Nozieres}. Eq.~\eqref{MNDLamb} was
obtained by writing down a generalized Dyson equation for the
electron Green's function in a constant window potential of width
$\hbar/\tau _{0}$, and solving it for all connected graphs in the
long-time limit. For this reason, the MND spectrum lacks formal
justification away from the threshold. In the present derivation,
we have taken into account the full perturbation at an arbitrary
time $t>0$, retaining only the first non adiabatic contribution in
the linked cluster expansion \cite{schotte}. We expect the effect
of higher order diagrams to be mainly concerned with the adiabatic
correction to the equilibrium energy and some additional
broadening of the excitation peaks. The latter should provide a
renormalization to the critical parameter. Nevertheless, in the
investigated ranges of temperatures and particle numbers, the
definition of $\alpha$ given here produces a markedly singular
response with the same range of criticality as the MND edge response
parameter~($\alpha =0-1$).

From this comparison with the free-gas case, we learn that the
trapping frequency $\omega$ enters crucially the physics of the
shake up process. Indeed, it modifies the long time response of
the gas as all single particle excitations  involve energy
exchanges which are now even multiples of $\hbar \omega$. This
gives rise to the periodic part of the fermion response and to the
corresponding spectral peaks which are then broadened at finite
temperatures due to the gaussian envelope, the latter being a
typical effect of suddenly switched perturbations
\cite{Anderson&Yuval}.
Up to now, we treated the response of the Fermi gas without any
reference to the dynamics of the impurity, that has just been
assumed in the excited states for $t\geq 0$. If, instead, the
two-level atom is subject to (say) a fast $\pi /2$ pulse and
quickly prepared in the superposition
$(\ket{g}+\ket{e})/\sqrt{2}$, it experiences a purely dephasing
dynamics due to the coupling with the gas, such that its state at
later times is $\rho_{\msc{imp}}(t)=( \ket{g} \bra{g} + \ket{e}
\bra{e} + \nu _{\beta }(t) \ket{g} \bra{e} + \mbox{h.c.})/2$. The
decoherence factor that enters the off-diagonal elements is just
the persistence amplitude that we obtained before, going to zero
at long times due to the orthogonality catastrophe. In the theory
of open systems, one typically uses a related function, the so
called Loschmidt echo $L(t)=|\nu_{\beta }(t)|^{2}$, which gives a
measure of the environmental response to the perturbation induced
by the system~\cite{diciotto,diciassette,zana} and which, as shown
in Ref.~\cite{pinja}, is linked with non-Markovianity of the open
system dynamics. The amount of non-Markovianity of a dynamical map
can be evaluated in different
manners~\cite{breuermeasure,rivas,fisherNM,geo}, which are,
however, essentially equivalent for a purely dephasing quantum
channel~\cite{altricin,pinja}. By adopting the definition in terms
of information flow given in~\cite{breuermeasure}, one finds
\begin{equation}
\mathcal{N}=\sum_{n} \left \{\left |\nu_{\beta}(t_{{\rm
max},n})\right |-\left | \nu_{\beta}(t_{{\rm min},n}) \right |
\right \}, \label{Ndef}
\end{equation}
where the sum is performed over all maxima and minima of the
$|\nu_{\beta}(t)|$, occurring at $t_{{\rm max},n}$ and $t_{{\rm
min},n}$, respectively. Using our previous results for the
amplitude, we can then obtain the non-Markovianity of the dynamics
of a two-level system in a trapped Fermi environment. The results
are shown in Fig.~\ref{nonmark}\textbf{B}, where we see that
$\mathcal{N}$ depends on the temperature and on the critical
parameter $\alpha$. In particular, it has a maximum at small
$\alpha$, increasing with low temperatures, and goes to zero both
for large temperatures~(due to the fact that thermal fluctuations
suppress oscillations in the persistence amplitude) and for
$\alpha >1$. In the latter case, excitations are generated at
every energy scale in the fermion gas, as witnessed by the fact
that the spectrum becomes structure-less. This implies that the
gas becomes more and more stiff~(in the sense that it is not able
to react on the impurity any more) and explains why $\mathcal{N}$
is zero: the open system does not receive information back, its
Loschmidt echo decays monotonously and thus the dynamics is
Markovian. As a result, we conclude that a non-Markovian dynamics
can be characterized, in our case, by the appearance of specific
spectral features in the excitation energy
distribution~\cite{nori12}.

We conclude with two remarks. First, the spectral distribution of
energy excitations obtained in the present work coincides with the
so called work distribution function, which is a central quantity
in non-equilibrium processes~\cite{campisi}. In the set-up that we
have described above, it is simple enough to conceive a `reverse'
protocol, with the fermi gas brought to thermal equilibrium in the
presence of the impurity~(i.e. with the two-level atom in the
excited state) which is then switched off. The comparison of the
work distribution functions in the direct and reverse protocols
would lead to a direct experimental test of the Crooks relation in
the quantum regime~\cite{crook}. The second remark is on the
experimental realization of the model that we have described. Many
experiments have recently dealt with the effects of impurities in
trapped Fermi gas~\cite{fermionexp}, and state-dependent
scattering lengths have been discussed~\cite{theo}. This would
lead to a direct test of our theory. Another viable candidate
could be a gas of hard-core bosons in one-dimension, where the
Loschmidt echo is equivalent to that of the corresponding Fermi
gas~\cite{johnpap} and in which impurities have recently been
experimentally generated~\cite{exp1d}.

\appendix

\section{Impurity potential}\label{AppA}
As explained in the main text, the non-interacting fermions in the
harmonic trap lie in their equilibrium configuration, set by
$\hat{H}_{0}$, until the impurity is excited and the sudden
perturbation $\hat{V}(t)=\hat{V}\theta (t)$ is felt by the gas. To
mimic a very strong difference in scattering length depending on
the internal state $\ket{e}$ of the impurity, we have modelled it
by a spatially localized potential, activated by the population of
the excited state, with the structure-less form  $V(x)=\pi
V_{0}x_{0}\delta (x)$.
\begin{figure}[b]
\centerline{\scalebox{0.95}{\includegraphics{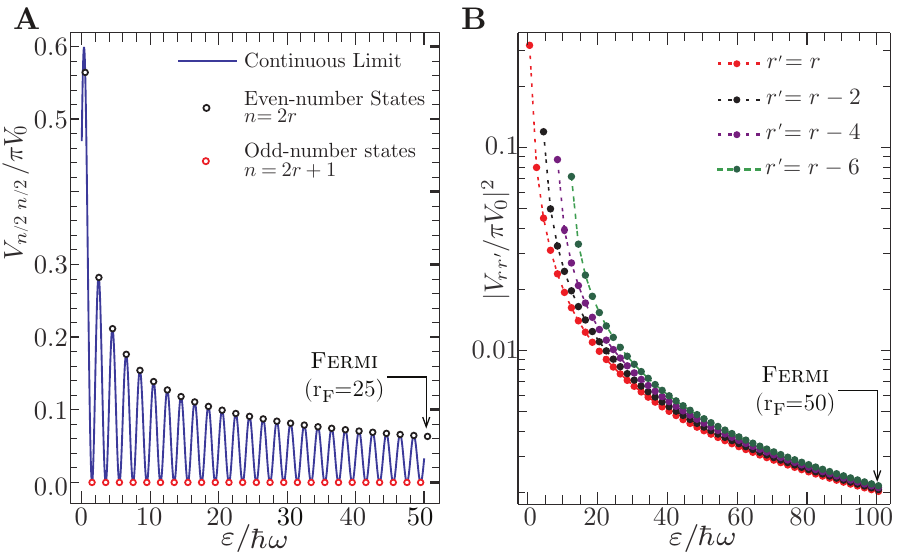}}}
\vskip -12pt \caption{(color on line) First order energy shifts
for the occupied fermion levels $\en$ of the trap~(\textbf{A}) and
second order couplings between even one-fermion
states~(\textbf{B}). The continuous limit representation of
$V_{n/2\,n/2}$ in panel~\textbf{A} is obtained by letting $n$ take
any real value between $0$ and $2\rF$, which corresponds to the
approximation $\eF/\hw \gg 1$. \Exclude{The terms in
panel~\textbf{B} enter the two-vertex graph~(\ref{Lambda2})}
\label{figuno} } \vskip -16pt
\end{figure}
For mathematical simplicity we have placed the impurity at the
minimum of the harmonic potential. Thus, the coupling matrix
elements between two unperturbed one-fermion states,
\begin{equation*}
\int{dx} \psi_{n}^{\ast}(x) V(x) \psi_{\np}(x) = \pi V_{0} x_{0}
\psi_{n}^{\ast }(0) \psi _{\np}(0)\text{,}
\end{equation*}
involve the Harmonic oscillator wave-functions at $x=0$. These
have the usual expression
\begin{equation*}
\psi_{n}(x) = \frac{x_{0}^{-1/2} \pi^{-1/4}}{2^{n/2}
n!^{1/2}}{\,}H_{n}\!\!\left(\frac{x}{x_{0}}\right){\,}e^{-\frac{x^2}{2x_{0}^2}}\text{,}
\end{equation*}
in terms of the Hermite polynomials $H_{n}$ with
$x_{0}\Exclude{=(m\omega/\hbar)^{-1/2}}$ being the characteristic
oscillator length. By the parity of the Hermite polynomials,
$H_{n}(-x/x_{0})=(-1)^{n}H_{n}(x/x_{0})$, we have $H_{n}(0)=0$ for
odd-$n$, i.e., $n=2r+1$, with $r=0,1,\cdots ,\infty $. Therefore,
the impurity potential induces excitations which connect only
unperturbed one-fermion states labelled by even numbers $n=2r$,
with $r=0,1,\cdots ,\infty $. The representation of the Hermite
polynomials with even numbers in power series,
\begin{equation*}
H_{2r}\!\!\left(\frac{x}{x_{0}}\right)=\sum_{k=0}^{r}\frac{4^{k}(2r)!{\,}(-1)^{r-k}}{(2k)!{\,}(r-k)!}\left(\frac{x}{x_{0}}\right)^{2k}\text{,}
\end{equation*}
allow us to write $H_{2r}(0)=(-1)^{r}\frac{(2r)!}{r!}$, which
leads to the matrix elements
\begin{align*}
V_{r\rp}=&\int{dx} \psi_{2r}^{\ast}(x) V(x) \psi_{2\rp}(x) \\
             =&\sqrt{\pi}V_{0}{\,}\frac{(-1)^{r+\rp}}{2^{r+\rp}}{\,}\frac{(2r)!^{1/2}}{r!}\frac{(2\rp)!^{1/2}}{\rp!},
\end{align*}
appearing in the second quantized representation of the impurity
potential in the harmonic oscillator basis. Using the identity
\begin{equation*}
\gamma_r = \frac{\Gamma (r+1/2)}{\Gamma (r+1)}\Exclude{
=(r+1)_{-1/2}}=2^{-2r} \pi^{1/2} \frac{(2r)!}{r!^2}\text{,}
\end{equation*}
where $\Gamma$ is the Euler gamma function \Exclude{and
$(r+1)_{-1/2}$ the  Pochhammer symbol}, we get
$V_{r\rp}=V_{0}(-1)^{r+\rp}{\,}{\gamma_{r}^{1/2}}{\,}{\gamma_{\rp}^{1/2}}$.
The diagonal matrix elements $V_{rr}=V_{0}\gamma_{r}$, shown in
Fig.~\ref{figuno}\textbf{A}, entirely determine the one-vertex
graph of Eq.~(\ref{Lambda1}), representing first order corrections
to the single-particle energies $\er=\hw (2r+1/2)$. On the other
hand, both diagonal and off-diagonal matrix elements of $V(x)$,
shown in Fig.~\ref{figuno}\textbf{B}, appear as absolute squares,
$|V_{r\rp}|^{2}=V_{0}^{2}\,\gamma _{r}\gamma _{\rp}$, in the
two-vertex graph given in Eq.~(\ref{Lambda2}).

\section{Linked Cluster expansion of the vacuum persistence amplitude}

\noindent The vacuum persistence amplitude, introduced in
Eqs.~(\ref{Eq:VacAmp}) and~\eqref{IP}, may be expanded by the
Dyson-Wick series
\begin{equation*}
\nu_{\beta }(t)=1+\sum_{m=1}^{\infty }\nu _{\beta }^{(m)}(t),
\end{equation*}
whose coefficients account for processes where the gas retrieves
its equilibrium unperturbed configuration after $m=1,2,\cdots$
`scatterings' with the impurity potential:
\begin{equation*}
\nu _{\beta }^{(m)}(t)=\frac{(-i)^m}{\hbar^{m}m!}
\int_{0}^{t}dt_{1} \cdots \int_{0}^{t}dt_{m} \big\langle
T\tilde{V}(t_{1})\cdots \tilde{V}(t_{m}) \big\rangle.
\end{equation*}
In the interaction picture,  with
\begin{eqnarray*}
\tilde{V}(t)=\sum_{r,\rp,\xi}V_{r\rp}\hat{c}_{2r\xi}^{\dg}(t)\hat{c}_{2\rp\xi}(t),
\end{eqnarray*}
 the time-evolution of creation and annihilation operators is that of the undisturbed Harmonic oscillator, i.e.,
\begin{equation*}
\hat{c}^{\dg}_{2r\xi}(t)=e^{i\omega \left(2r+\frac{1}{2}\right) t
}\hat{c}^{\dg}_{2r\xi}, \quad \hat{c}_{2r\xi}(t)=e^{-i\omega
\left(2r+\frac{1}{2}\right) t}\hat{c}_{2r\xi}.
\end{equation*}
Accordingly,
\begin{align*}
& \big\langle T\tilde{V}(t_{1})\cdots \tilde{V}(t_{m})\big\rangle =\sum_{r_{1},r_{1}^{\p},\xi _{1}}V_{r_{1}r_{1}^{\p}}\cdots \sum_{r_{m},r_{m}^{\p},\xi _{m}}V_{r_{m}r_{m}^{\p}} \\
& \qquad \times \big\langle T\hat{c}_{2r_{1}\xi _{1}}^{\dg
}(t_{1})\hat{c}_{2r_{1}^{\p}\xi _{1}}(t_{1})\cdots
\hat{c}_{2r_{m}\xi _{m}}^{\dg }(t_{m})\hat{c}_{2r_{m}^{\p}\xi
_{m}}(t_{m})\big\rangle.
\end{align*}
Here, the time-ordered average  at the right-hand side may be
decomposed using  the Wick's theorem into sums of products where
each factor is a contracted pairs of creation/anihilation
operators. The central approximation of the work is to retain
terms that contain only equal time and two-time contractions,
namely
\begin{equation*}
\big\langle {\cdots} \Include{ \contraction{} {\hat{c}} {_{2 r_{j}
\xi_{j}}^{\dg}(t_{j})} {\hat{c}} } \hat{c}_{2 r_{j}
\xi_{j}}^{\dg}(t_{j}) \hat{c}_{2r_{j}^{\p}\xi_j}(t_{j})
{\cdots}\big\rangle={\cdots} f_{r_{j}}^{+} \delta
_{r_{j}\,r_{j}^{\p}} {\cdots}
\end{equation*}
and
\begin{align*}
&\big\langle {\cdots} \Include{ \bcontraction{} {\hat{c}}
{_{2r_{j}\xi_j}^{\dg}(t_{j})\hat{c}_{2r_{j}^{\p}\xi_j}(t_{j}){\cdots}\hat{c}_{2r_{i}\xi}^{\dg}(t_{i})}
{\hat{c}_{2}}
\acontraction{{\:\:}\hat{c}_{2r_{j}\xi}^{\dg}(t_{j})} {\hat{c}}
{_{2r_{j}^{\p}\xi_j}(t_{j}} {){\cdots}\hat{c}_{2r\xi}} }
\hat{c}_{2r_{j}\xi_j}^{\dg}(t_{j})\hat{c}_{2r_{j}^{\p}\xi_j}(t_{j})
{\cdots}
\hat{c}_{2r_{i}\xi_i}^{\dg}(t_{i})\hat{c}_{2r_{i}^{\p}\xi_i}(t_{i})
{\cdots}\big\rangle
\\
& \qquad =-\hbar ^{2} {\cdots} \delta_{\xi_i \xi_j}
G_{r_{j}}^{\beta}(t_{i}-t_{j}) \delta_{r_{j}r_{i}^{\p}}
G_{r_{i}}^{\beta }(t_{j}-t_{i}) \delta_{r_{i}r_{j}^{\p}} {\cdots},
\end{align*}
where the Fermion occupation numbers $f_{r}^{+}$ and  the
unperturbed Fermion propagators $G_{r}^{\beta }(t)$ have been
introduced in the main text~(see Eq.~(\ref{prop})). When these
expressions are summed over even level numbers and integrated over
time variables, we are left with products including either
\begin{equation*}
\Lambda _{1}^{\beta }(t)=-\frac{i\spind t}{\hbar }
\sum_{r}V_{rr}f_{r}^{+}
\end{equation*}
or
\begin{align*}
& \Lambda _{2}^{\beta }(t)=-\frac{\spind }{2}\sum_{\rp,\rs}|V_{\rp\rs}|^{2} \notag \\
& \qquad \qquad \times \int_{0}^{t}d\tp\int_{0}^{t}d\ts
G_{\rp}^{\beta }(\tp-\ts)G_{\rs}^{\beta }(\ts-\tp),
\end{align*}
which are just the connected diagrams reported in
Eqs.~(\ref{Lambda1}) and~(\ref{Lambda2}). Each product equals
$\Lambda _{1}^{\beta }(t)^{j} \Lambda _{2}^{\beta }(t)^{m-j}$,
obtained by $\binom{m}{j}$ distinct contractions for some $j$
between $0$ and $m$.

\noindent This means that
\begin{equation*}
\nu _{\beta }^{(m)}(t)\approx \frac{1}{m!} \sum_{j=0}^{m}
\binom{m}{j} \Lambda_{1}^{\beta }(t)^{j} \Lambda _{2}^{\beta
}(t)^{m-j}.
\end{equation*}
so that the Dyson-Wick series for the vacuum persistence amplitude
takes the exponential form
\begin{equation*}
\nu _{\beta }(t)=\sum_{m=0}^{\infty }\frac{[\Lambda _{1}^{\beta
}(t)+\Lambda _{2}^{\beta }(t)]^{m}}{m!}=e^{\Lambda _{1}^{\beta
}(t)+\Lambda _{2}^{\beta }(t)}.
\end{equation*}
The single vertex graph~(\ref{Lambda1}) gives rise to the first
order energy shift discussed in the main text, while the
two-vertex connected graph has a more involved structure.
Performing the time-ordered integrals in~(\ref{Lambda2}) we
rewrite it  as
\begin{equation}
\Lambda_{2}^{\beta }(t)=-\frac{\alpha
\eF}{\hbar}\sum_{r,\rp=0}^{\infty}
\left[it\varphi_{r\rp}+\frac{\psi _{r\rp}(t)}{2\omega }\right]
f^{+}_{r} f^{-}_{\rp}, \label{L2Start}
\end{equation}
in which
\begin{equation*}
\varphi_{r\rp}=\frac{\gamma_r{\,}\gamma_{\rp}}{r-\rp}, \qquad
\psi_{r\rp}(t)=\varphi_{r\rp}{\,}\frac{1-e^{2 i (r-\rp) t
\omega}}{r-\rp}.
\end{equation*}
Here, we may separate the sums over even-state labels~($r$,$\rp$),
so that the two-vertex connected graph can then be split into
three contributions:
\begin{equation*}
\Lambda _{2}^{\beta}(t) = \Lambda_{2\msc{s}}^{\beta }(t) +
\Lambda_{2\msc{g}}^{\beta }(t) + \Lambda_{2\msc{p}}^{\beta }(t),
\end{equation*}
where the subscripts stand for Shift, Gaussian and Periodic,
respectively. In particular:
\begin{itemize}
\item[\textbf{(i) }] the off-diagonal summands in~(\ref{L2Start})
that multiply $\varphi_{r\rp}$ give rise to
$\Lambda_{2\msc{s}}^{\beta}(t)=-itE_{2}^{\beta }/\hbar$, where
\begin{equation}
E_{2}^{\beta } =\alpha \eF \sum\limits_{r{\neq}\rp=0}^{\infty}
f^{+}_{r}\,\varphi_{r\rp}\,f^{-}_{\rp}.
\end{equation}
is the energy correction reported in Eq.~\eqref{Delta2B}
\item[\textbf{(ii) }] the diagonal elements of Eq.~(\ref{L2Start})
yield the quadratic power law in  $\Lambda_{2G}^{\beta}(t)$~[see
Eq.~\eqref{eq:gbeta}];

\item[\textbf{(iii) }] the remaining terms of the series in
Eq.~(\ref{L2Start}) give the time periodic sub-diagram
\begin{equation}
\Lambda _{2\msc{p}}^{\beta }(t)=-\frac{\alpha
\eF}{2\hw}\sum\limits_{r{\neq
}\rp}^{\infty}\,f_{r}^{+}\,\psi_{r\rp}\,f_{\rp}^{-},
\end{equation}
also reported in Eq.~(\ref{Lambda2E}).
\end{itemize}

\section{Numerical computations}\label{AppC}

As shown in the main text, the real and imaginary parts of the
connected graphs $\Lambda _{1}^{\beta }(t)$~[Eqs.~\eqref{Lambda1},
Fig.~\ref{eradue}\textbf{B}] and $\Lambda _{2}^{\beta
}(t)$~[Eq.~\eqref{Lambda2},  Fig.~\ref{eradue}\textbf{C}, Fig.~2]
combine in the vacuum persistence amplitude to give:
\begin{equation*}
\nu _{\beta }(t)= e^{-\frac{i t}{\hbar}(E_{1}^{\beta
}+E_{2}^{\beta })}\,
e^{-\alpha\,g_{\beta}\,\omega^{2}t^{2}}\,e^{\Lambda
_{2\msc{p}}^{\beta }(t)}.
\end{equation*}
The knowledge of $\nu _{\beta }(t)$, allows us to determine the
decoherence factor $|\nu _{\beta
}(t)|$~(Fig.~\ref{cinque}~\textbf{A}, \textbf{C}, \textbf{E}), the
shake up spectrum $|\tilde{\nu} _{\beta
}(E)|$~(Fig.~\ref{cinque}\textbf{B}, \textbf{D}, \textbf{F},
Fig~\ref{nonmark}\textbf{A}), and the non-Markovianity measure
$\mathcal{N}$~[Eq.~\eqref{Ndef}, Fig~\ref{nonmark}\textbf{B}].
Then, the basic quantities in our calculations are:
\begin{itemize}
\item[\textbf{(i) }] the first and second order corrections,
$E_{1}^{\beta}$~[Eq.~(\ref{eq:E1}), Fig.~\ref{eradue}\textbf{B}]
and $E_{2}^{\beta}$~[Eqs.~(\ref{Delta2B}),
Fig.~\ref{eradue}\textbf{C}], to the equilibrium
energy~$E_{0}^{\beta }$~ (Fig.~\ref{eradue}\textbf{A});
 \Exclude{Recall that
$E_{0}^{\beta }$ is plotted in Fig.~\ref{eradue}\textbf{A}, while
$E_{1}^{\beta }$ and $E_{2}^{\beta }$, perturbation corrections,
due to the impurity potential are shown in
Figs.~\ref{eradue}\textbf{B} and~\textbf{C} of the main text;}
\item[\textbf{(ii) }] the coefficient $g_{\beta }$ determining the
standard deviation $\delta_{\beta}$~[Eq.~(\ref{eq:gbeta}),
Fig.~\ref{eratre}\textbf{A}] of the Gaussian sub-diagram $\Lambda
_{2\msc{g}}^{\beta }$; \item[\textbf{(iii) }] the shake-up
sub-diagram $\Lambda_{2\msc{p}}^{\beta}$~[Eqs.~(\ref{Lambda2E}),
Fig.~\ref{eratre}\textbf{B}].
\end{itemize}

\begin{figure}[!!h]
\centering \scalebox{0.78}{\includegraphics{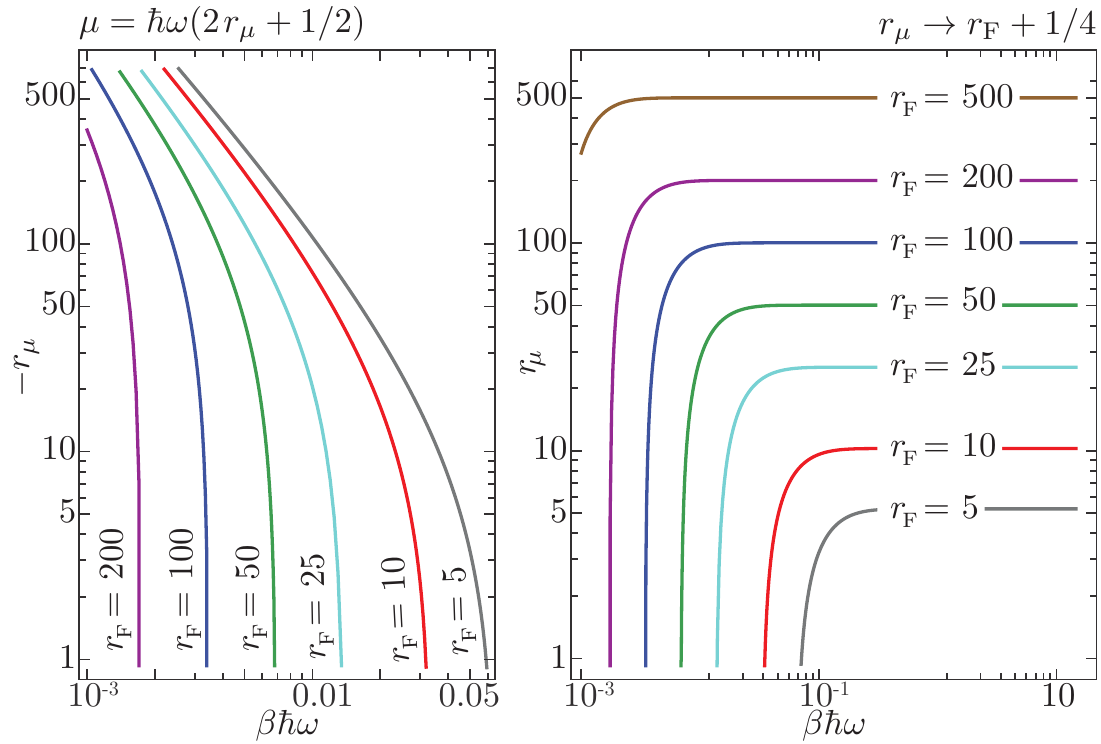}}
\caption{Chemical potential index $r_{\mu}$ vs $\beta\hw$, for
$\rF=5-500$} \label{figMmu}
\end{figure}

These contributions contain summations running over all
one-fermion eigenstates of the trap weighted by the fermi factors
$f_{r}^{+}$, $f_{\rp}^{-}$. These are expressed as
$f^{+}_{r}=[1+e^{2 \beta\hw (r-r_{\mu})}]^{-1}$ and
$f^{-}_{\rp}=[1+e^{-2 \beta\hw (r-r_{\mu})}]^{-1}$, using the
parametrization $\mu =\hw(2r_{\mu }+1/2)$ for the chemical
potential. The index $r_{\mu }$ tends to $\rF + 1/4$ for $\beta
\to \infty$, so that the $\mu$ lies in the middle between the
highest occupied~($\eF=\varepsilon_{2\rF}$) and the lowest
unoccupied one-fermion levels~($\varepsilon_{2\rF+1}$) of the gas.
For finite $\beta $ we determined $r_{\mu}$ by numerically
constraining the conservation of particle number:
\begin{equation*}
2\rF+1=\sum_{r}\left(f_{r}^{+}+f_{r+1/2}^{+}\right),
\end{equation*}
As shown in Fig.~\ref{figMmu}, $r_{\mu}$ and hence $\mu$, reach
their maximum values at the absolute zero~($\beta \to \infty $).
They decrease with decreasing $\beta$ and take largely negative
values for  $\beta \to 0$ where the classical limit applies.
Interestingly enough, both $r_{\mu}$ and  $\mu$ are almost
independent on temperature for $\beta\hw \gtrsim 0.4$ and
$\rF=5-500$.

\begin{figure}[!!h]
\centerline{\scalebox{0.95}{\includegraphics{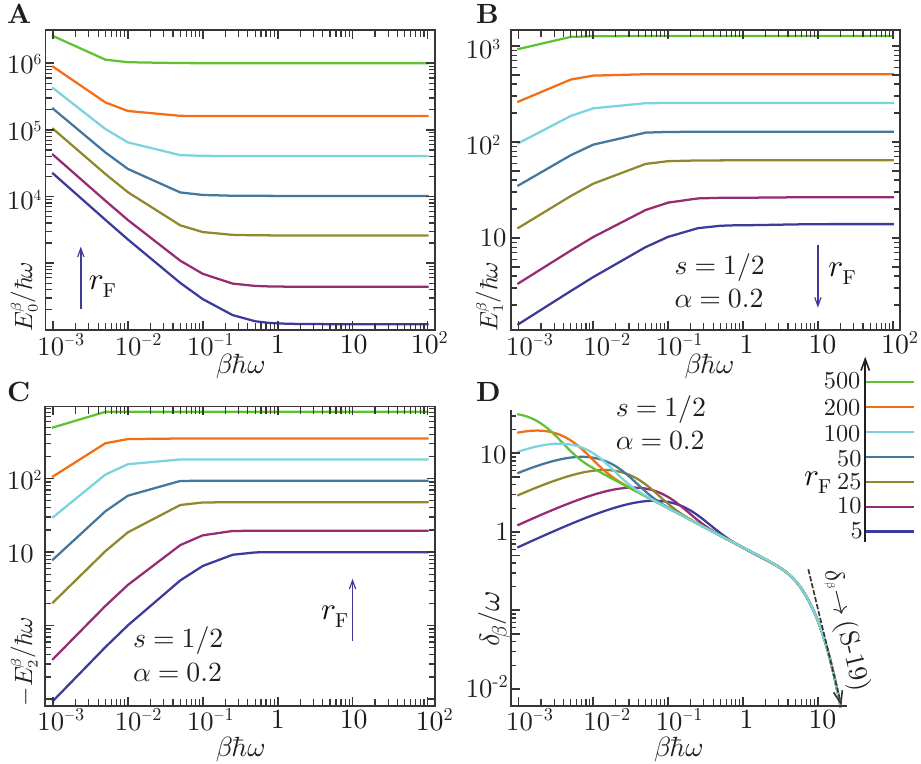}}}
\vskip -8pt \caption{ (color on line) Equilibrium energy
$E_{\beta}^{0}$ of a spin-$1/2$ gas into a harmonic
trap~(panel~\textbf{A}) compared with first and second order
corrections, $E_{\beta}^{1}$~[Eq.~(\ref{eq:E1}), panel~\textbf{B}]
and $E_{\beta}^{2}$~[Eq.~(\ref{Delta2B}), panel~\textbf{C}], due
to the sudden impurity potential $V(x)$ with $\alpha
=0.1,0.4$~(see also Fig.~1 of the main text). All Energy values
are expressed in units of $\hw$ \textit{vs} $\beta\hw$ for
different Fermi numbers/energies and temperatures. Significant
changes are induced by $\rF$, whereas negligible differences are
observed in the energy curves for $\beta\hw > 0.4$. Gaussian
standard deviation $\delta_{\beta}$~(panel~\textbf{D}) appearing
in the two-vertex contribution $\Lambda_{2\msc{g}}^{\beta }$ and
reported in units of $\omega$ for the same values of $\rF$ and
$\alpha$ used in the plots of panels (\textbf{A}-\textbf{C}).
\label{FigE} }
\end{figure}

With the computed $f^{\pm}_{r}$-distributions, we run numerical
computations of the basic quantities (i)-(iii) using a high energy
cut-off $\varepsilon_{\msc{cut}}\gg \eF$. To fix
$\varepsilon_{\msc{cut}}= \hw (2r_{\msc{cut}}+1/2)$, we performed
convergency tests by changing $r_{\msc{cut}}$ in order to have a
maximum instability error below $0.1\%$.
\\
In our applications to a spin~$1/2$ gas, we observed that accurate
estimations of $E_{2}^{\beta }$ and $g_{\beta }$ require values of
$r_{\msc{cut}}$ of the order of $10^{4}$ for $\rF$ below $\sim
200$ and $\beta\hw $ larger than $\sim 10^{-5}$. In particular,
the plots of Fig.~1\textbf{C} and Fig.~2\textbf{A} of the main
text were generated with a energy cut-off of $10-10^3~\eF$. As for
$E_{0}^{\beta }$, $E_{1}^{\beta }$, and $\Lambda_{2\msc{p}}^{\beta
}$, we found out a cut-off of $\sim 10~\eF$ to be sufficient in
the investigated ranges of fermion numbers\Exclude{, critical
exponents} and temperatures. Accordingly, we included up to
$10^{3}$ one-fermion states in Fig.~1\textbf{A}-\textbf{B} and
Fig.~2\textbf{C} of the main text.
\\
To provide a more complete picture of the basic quantities
involved in the numerical calculations,  in
Fig.~\ref{FigE}\textbf{A}-\textbf{C} we report the behavior of the
equilibrium energy and the two energy shifts  \textit{vs} $\beta
\hw$ for gases made of different fermion numbers. We remark that
both $E_{1}^{\beta }$ and $E_{2}^{\beta }$ are indeed small
corrections to the unperturbed value $E_{0}^{\beta }$ of a
spin-$1/2$ gas, for values of the critical exponent $\alpha
\lesssim 1$. In addition, $E_{1}^{\beta }$ is generally larger
that $E_{2}^{\beta }$. These energies are strongly affected by the
number of particles in the gas and weakly dependent on temperature
for $\beta\hw \gtrsim 0.4$.
\\
On the other hand the Gaussian damping/broadening brought by
$\Lambda_{2\msc{g}}^{\beta }$, with standard deviation
$\delta_{\beta}$, is almost entirely dependent on the thermal
energy $\beta\hw$ and the critical exponent
$\alpha$~(Fig.~\ref{FigE}\textbf{D}). This contribution leads to a
smearing of the shake up response of the system in way that
resembles the Anderson-Yuval approach to the Kondo
problem~\cite{Anderson&Yuval}. Indeed, $\delta_{\beta}$ decreases
exponentially to zero with increasing $\beta\hw$, following the
limiting trend
\begin{equation}
\delta _{\beta }{\approx }2^{3/2}\alpha ^{1/2} e^{-\beta\omega
\hbar/4}
\end{equation}
for $\beta\hw \gtrsim 7-8$~(Figs.~2\textbf{A}
and~\ref{FigE}\textbf{B}).

\begin{figure}[h]
\centerline{\scalebox{0.99}{\includegraphics{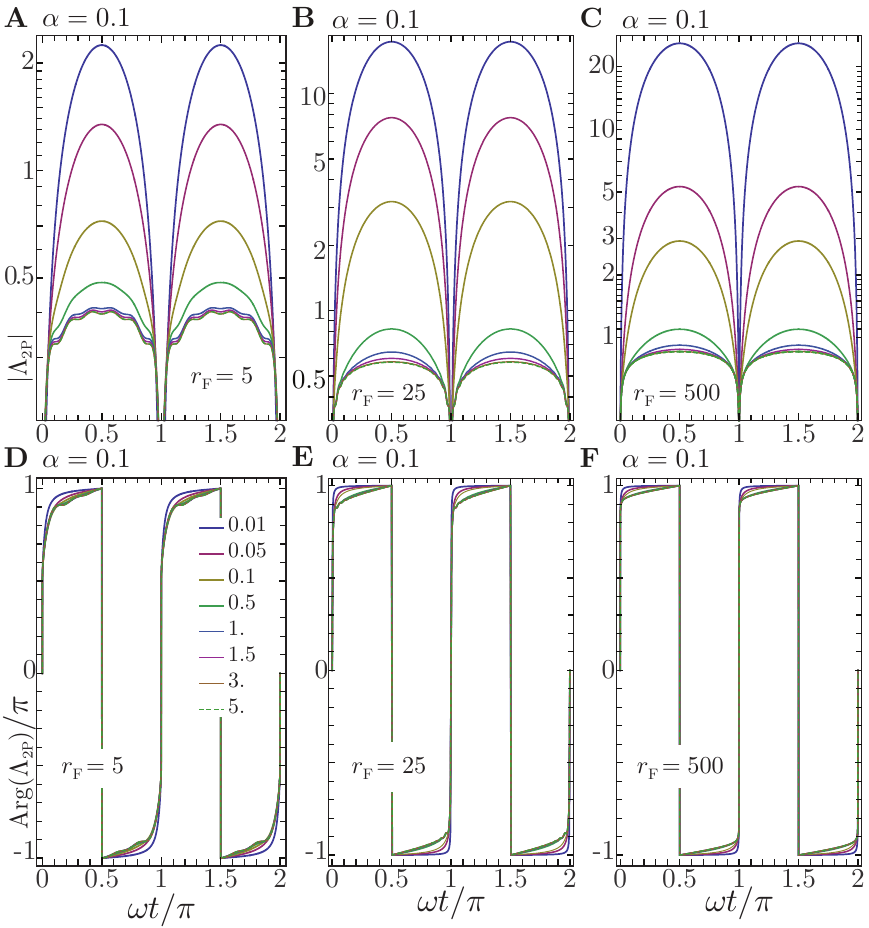}}}
\vskip-4pt \caption{(color online): Modulus and phase of the
periodic component $\Lambda_{2\msc{p}}^{\beta }$~(\ref{Lambda2E})
\textit{vs} $\omega t$, for $\beta\hw = 0.01-5$ and $\alpha =
0.1$. Several fermion numbers are tested, i.e., $\rF=5$
(\textbf{A},\textbf{D}), $\rF=25$ (\textbf{B},\textbf{E}), and
$\rF=500$ (\textbf{C},\textbf{F}). \label{FigGP}} \vskip -4pt
\end{figure}

In addition, as evident by comparing Fig.~2\textbf{B} of the main
text with the plots of Fig.~\ref{FigGP}, the sub-diagram
$\Lambda_{2\msc{p}}^{\beta }$ has a time period of $\pi/\omega$.
Its modulus $|\Lambda_{2\msc{p}}^{\beta }|$ presents zeroes at
$\omega t = m \pi$ and maxima at $\omega t = m \pi/2$, with
$m=0,\pm 1,\pm 2,\cdots$. The intensities of such
maxima~(Fig.~\ref{FigGP}\textbf{A}-\textbf{C}) increase with
increasing the Fermi number~($2\rF$), the critical
exponent~($\alpha$), and the thermal energy~($\beta^{-1}$). The
phase of $\Lambda_{2\msc{p}}^{\beta }$ is discontinuous at the
extremes of $|\Lambda_{2\msc{p}}^{\beta }|$ and less dependent on
these parameters~(Fig.~\ref{FigGP}\textbf{D}-\textbf{F}).

\section{Low thermal energy approximation}

We work at sufficiently low temperatures such that the chemical
potential is well approximated by its maximum value
\begin{equation}
\label{Mmu} \mu \underset{\beta\hw\rightarrow \infty }{\rightarrow
}\hw(2\rF+1),
\end{equation}
corresponding to
\begin{equation}
\label{Rmu} r_{\mu }\underset{\beta\hw\rightarrow \infty
}{\rightarrow }\rF+1/4,
\end{equation}
as shown in Fig.~\ref{figMmu}. Then, we consider systems with a
relatively large number of
particles~($\rF\gtrsim 10$) and focus on the sub-diagrams $\Lambda _{2\msc{g}%
}^{\beta }(t)$ and $\Lambda _{2\msc{p}}^{\beta }(t)$. In this, way
we provide the formal justifications for the analytical
approximations introduced in Eqs.~(\ref{Lambda2GA})
and~(\ref{LambXX}), which determine the `unshifted' amplitude $\nu
_{\beta }^{\p}(t)$ given in Eq.~(\ref{Vpbet}) and the excitation
spectrum $\tilde{\nu}_{\beta }^{\p}(E)$, shown in
Fig~\ref{nonmark}\textbf{A}.

\begin{figure}[h]
\scalebox{0.99}{\includegraphics{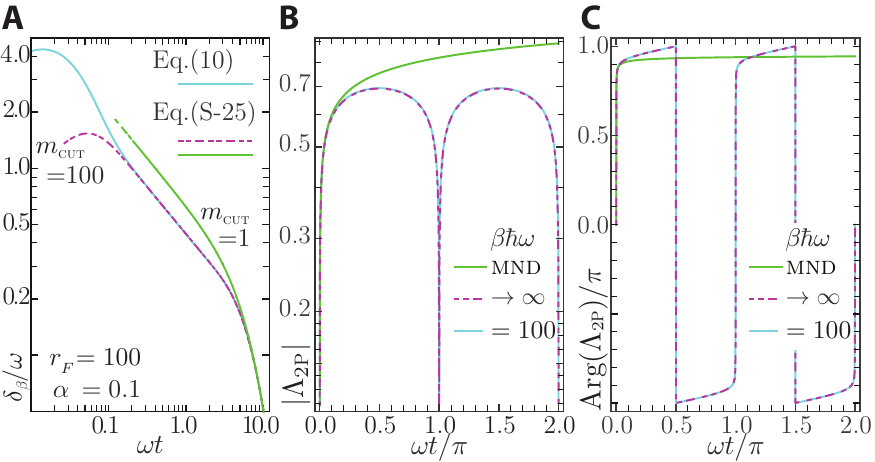}} \caption{(color
online) (\textbf{A}) Standard deviation $\delta_{\beta }$ of the
Gaussian diagram $\Lambda _{2\msc{g}}^{\beta}(t)$;
(\textbf{B},\textbf{C}) absolute value and phase of the Fermi edge
diagram $\Lambda _{2\msc{p}}^{\beta }(t)$, for $\protect\alpha
=0.1$ and $\rF=100$. In panel~\textbf{A}, numerical computations
from Eq.~(\ref{eq:gbeta}) are compared with the truncated
series~\eqref{deltabAPP} for $m_{\msc{cut}}=1,100$. In
panels~\textbf{B},\textbf{C}, numerical computations of $\Lambda
_{2\msc{p}}^{\protect\beta }$, for $\beta \hw=100$, are shown
together with the analytical approximation obtained from
Eq.~(\protect\ref{LambFES}), in which $\protect\omega \protect\tau
_{0}=0.001 $, and the results form the MND theory~[see
Eq.~(\protect\ref{MNDLamb})].} \label{LambdaPASX}
\end{figure}

\noindent Using the power series
\[
f_{r}^{+}f_{r}^{-}\underset{\er\lessgtr \mu
}{=}\sum\limits_{m=1}^{\infty }(-1)^{m+1}\,m\,e^{\pm \beta
m(\er-\mu )},
\]
we rewrite the Gaussian damping parameter, reported in
Eq.~(\ref{eq:gbeta}), as
\[
g_{\beta }=\sum_{m=1}^{\infty }(-1)^{m+1}\,m\,g_{m}^{\beta }.
\]%
The coefficients of this expansion read%
\begin{align*}
g_{m}^{\beta }& =\frac{\eF}{\hw}\sum_{r<r_{\mu}}\gamma _{r}^{2}e^{\beta m(\er%
-\mu )}+\frac{\eF}{\hw}\sum_{r>r_{\mu}}\gamma _{r}^{2}e^{-\beta m(\er%
-\mu )} \\
& =\frac{\eF}{\hw}\sum_{r<r_{\mu}}\gamma _{r}^{2}e^{-2\omega \tau
_{m}(r_{\mu }-r)}+\frac{\eF}{\hw}\sum_{r>r_{\mu}}\gamma
_{r}^{2}e^{-2\omega \tau _{m}(r-r_{\mu })},
\end{align*}
with $\tau _{m}$ being the characteristic times $\tau
_{m}=m\beta\hbar $, induced by thermal fluctuations. Then, we use
Eq.~(\ref{Rmu}) and perform a change of summation indices to write
\begin{eqnarray*}
g_{m}^{\beta } &=&\frac{\eF}{\hw}e^{-2\omega \tau _{m}(r_{\mu }-\rF%
)}\sum_{r=0}^{\rF}\gamma _{\rF-r}^{2}e^{-2\omega \tau _{m}r} \\
&&+\frac{\eF}{\hw}e^{2\omega \tau _{m}(r_{\mu
}-\rF)}\sum_{r=1}^{\infty }\gamma _{\rF+r}^{2}e^{-2\omega \tau
_{m}r}.
\end{eqnarray*}%
The transformed summations in this last line are dominated by low
$r$ terms.
In a many fermion environment, we may use the asymptotic relation $\gamma _{%
\rF\pm r}^{2}\approx \gamma _{\rF}^{2}\approx \rF^{-1}$ and obtain
\[
g_{m}^{\beta }\underset{\rF\gg1}{\approx }2\frac{e^{2\omega \tau
_{m}(r_{\mu
}-\rF)}+e^{2\omega \tau _{m}(\rF-r_{\mu }+1)}-e^{-2\omega \tau _{m}r_{\mu }}%
}{e^{2\omega \tau _{m}}-1}.
\]%
Then, using the asymptotic form~\eqref{Rmu} and neglecting
$e^{-2\omega \tau _{m}r_{\mu }}$, we find
\[
g_{m}^{\beta }\approx 2\frac{e^{\omega \tau _{m}/2}}{e^{\omega
\tau _{m}}-1},
\]%
which leads to Eq.~(\ref{Lambda2GA}), i.e.,%
\[
g_{\beta }\approx 2\sum_{m=1}^{\infty }(-1)^{m}m\frac{e^{\omega \tau _{m}/2}%
}{e^{\omega \tau _{m}}-1},
\]%
and let us approximate the standard deviation with
\begin{equation}
\delta _{\beta }\approx 2\alpha ^{1/2} \left[ \sum_{m=1}^{m_{\msc{cut}%
}}(-1)^{m}m\frac{e^{\omega \tau _{m}/2}}{e^{\omega \tau
_{m}}-1}\right] ^{1/2}.  \label{deltabAPP}
\end{equation}%
Eq.~(\ref{deltabAPP}), plotted in Fig.~\ref{LambdaPASX}\textbf{A}
for $m_{\msc{cut}}=1,100$, is independent of the number of
particles in the gas. We have verified that the truncated series
for $m_{\msc{cut}}=100$ works extremely well for $\rF=5-500$ and
$\beta\hw \gtrsim 0.4$. The asymptotic form of $\delta _{\beta }$
leads to the result reported in Fig2\textbf{A} of the main text.

\noindent As for the Fermi-edge component $\Lambda
_{2\msc{p}}^{\beta }(t)$, we consider the auxiliary functions
$\lambda _{\pm }^{\beta }(t)$ introduced
in the main text, which enter the connected graph $\Lambda _{2}^{\beta }(t)$%
~(see Eq.~\eqref{Lambda2}). We replace the particle-hole distributions $%
f_{r}^{\pm }$ with the power series expansion%
\begin{align*}
f_{r}^{\pm }=& \sum\limits_{m=0}^{\infty }(-1)^{m}e^{\pm \beta
m(\er-\mu )}
\qquad \quad \er \lessgtr \mu  \\
=& -\sum\limits_{m=1}^{\infty }(-1)^{m}e^{\mp \beta m(\er-\mu )}
\qquad \er\gtrless \mu
\end{align*}
to write
\begin{equation}
\lambda _{\pm }^{\beta }(t)=\sum_{m=0}^{\infty }(-1)^{m}\,\lambda
_{m\pm }^{\beta }(t).  \label{lambdasexp}
\end{equation}%
Here, the coefficients $\lambda _{m\pm }^{\beta }(t)$ may be
computed exactly by the finite summations
\[
\sum_{r=r_{1}}^{r_{2}}\gamma
_{r}\,z^{r}=z^{r_{1}}\,\HypFR(r_{1},z)-z^{r_{2}+1}\,\HypFR(r_{2}+1,z),
\]
holding for any $z\neq 1$, in which $\HypFR$ is the regularized
Hypergeometric function
\begin{equation}
\HypFR(\rF,z)=\gamma _{\rF}\;\HypF(1,1/2+\rF,1+\rF;z)
\end{equation}
Working in the temperature range where Eq.~\eqref{Mmu} holds, the
$m=0$
coefficients of the series~(\ref{lambdasexp}) turn out to be independent on $%
\beta $:
\begin{eqnarray}
\lambda _{0+}^{\beta }(t) &=&\frac{\sqrt{\pi
}}{\sqrt{1-e^{2it\omega }}}
\label{lambdasexpM} \\
&&-e^{\frac{it\eF}{\hbar }+\frac{3\omega
it}{2}}\,\HypFR(\rF+1,e^{2it\omega
}),\qquad \qquad  \notag \\
\lambda _{0-}^{\beta }(t) &=&e^{-\frac{it\eF}{\hbar }-\frac{3\omega it}{2}}\,%
\HypFR(\rF+1,e^{-2it\omega }). \label{lambdasexpP}
\end{eqnarray}
The other coefficients, accounting for low temperature effects,
have the form:
\begin{eqnarray*}
\lambda _{m\pm }^{\beta }(t) &=&\pm \frac{\sqrt{\pi }e^{-\beta m\eF}}{\sqrt{%
1-e^{2\omega \left( \tau _{m}\pm it\right) }}} \\
&&\mp e^{\pm \frac{it\eF}{\hbar }\pm \frac{3}{2}\omega \left(
it+\tau _{m}\right) }\HypFR\lbrack \rF+1,e^{2\omega \left( \tau
_{m}\pm it\right) }]
\nonumber \\
&&\mp e^{\pm \frac{it\eF}{\hbar }\pm \frac{3}{2}\omega \left(
it-\tau _{m}\right) }\HypFR\lbrack \rF+1,e^{2\omega \left( \pm
it-\tau _{m}\right) }],
\end{eqnarray*}
With the zero temperature parts~(\ref{lambdasexpP})
and~(\ref{lambdasexpM}), we need to add an imaginary time
regularization to the $\ts$-integral in the two-loop
term~(\ref{Lambda2}), i.e., we have to shift the $\ts$ integration
domain by $i\tau _{0}$ to prevent $\HypFR(\rF+1,e^{\pm 2\omega
ri\ts})$ from being singular. We may, then, insert
Eq.~(\ref{lambdasexp}) in the expression for $\Lambda _{2}^{\beta
}(t) $, compute the $r$-summations, and use the large $\rF$
expansion
\begin{equation*}
\HypFR(\rF\gg 1,z)=\frac{\rF^{-1/2}}{1-z}+\mathrm{o}\left(
\rF^{-3/2}\right)\text{.}
\end{equation*}
Finally, we may perform the $\tp$ and $\ts$-integrals, excluding
terms proportional to $t$ and $t^{2}$. What is left is a
combination of logarithmic and polylogarithmic functions,
dominated by the Fermi-edge term reported in Eq.~(\ref{LambXX}).
Indeed, as shown in Fig.~\ref{LambdaPASX}, the zero temperature
form
\begin{equation}
\Lambda _{2\msc{p}}^{\infty }(t)\approx \ln \left[ \frac{e^{2\tau
_{0}\omega }-1}{e^{2\omega (\tau _{0}+it)}-1}\right] ^{\alpha }
\label{LambFES}
\end{equation}
obtained from this procedure is in excellent agreement with the
numerical
calculations reported in Fig.~\ref{eratre}\textbf{B} for $\rF=100$ and $%
\beta\hw$ larger than $\sim 1$. The reliability of such an
approximation is also attested by the comparison in
Fig~\ref{nonmark}\textbf{A}. We will see in the following appendix
that Eq.~\eqref{LambFES} provides an accurate description of the
singular response of a Fermi gas with low numbers of particles.

In the main text, we observed that $\Lambda _{2\msc{p}}^{\infty
}(t)$ correctly tends to the MND form given by Eq.~(\ref{MNDLamb})
when the harmonic trap frequency is lowered to zero, keeping the
number of particles in the gas finite. The singularity index is
proportional to the height of the impurity potential barrier.
Nozi\`{e}res and De~Dominicis~\cite{Nozieres} calculated the
propagator for a free electron in the transient potential
activated by a structure-less core-hole, by solving the associated
Dyson equation in the long-time limit with Muskhelishvili
techniques for singular integral equations. They assumed a
constant potential of arbitrary height and width $\hbar/\tau_0$
around the Fermi level of the gas and found the singularity index
to depend of the phase-shifts of this potential. Therefore they
provided an asymptotic expression for all closed loops
$\Lambda^{\infty
}(t)=\sum_n \Lambda_n^{\infty }(t)$, which may be written as Eq.~%
\eqref{MNDLamb}. In our derivation, we have given a model to the
impurity potential, being described by non constant matrix
elements $V_{r\rp}$~(see appendix~\ref{AppA}). In addition we have
found an analytical form for two
vertex graph, Eq.~(\ref{LambXX}), which accurately describes $\Lambda _{2%
\msc{p}}^{\beta}(t)$ at any time $t$ for a sufficiently wide range
of temperatures and particle number. We expect the non trivial
part of the higher order contributions $\Lambda _{n>2}^{\beta}(t)$
to change the value of the $\alpha$-parameter in such a way that
it will depend on the phase shifts of the impurity potential.

\begin{figure}[h]
\scalebox{0.95}{\includegraphics{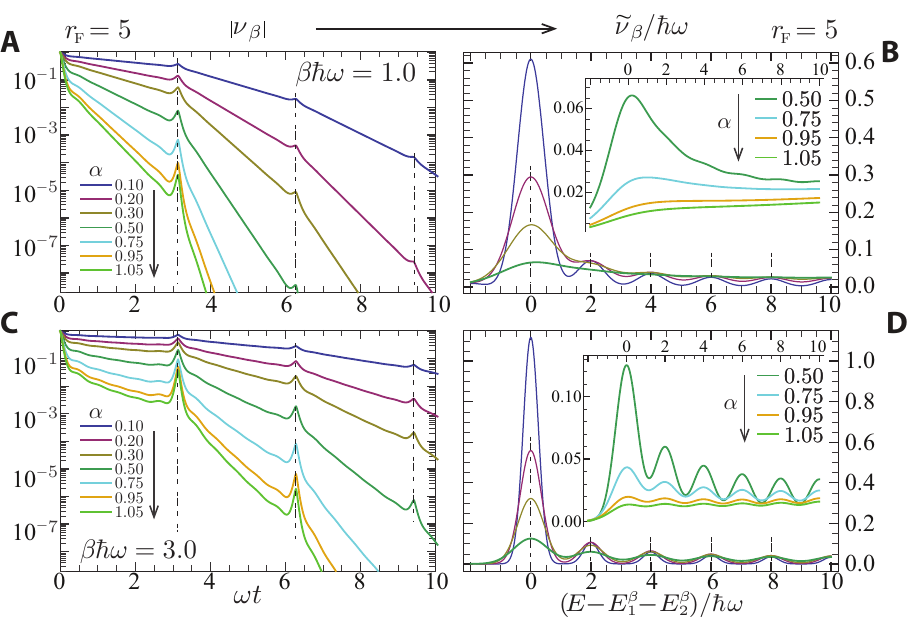}} \vskip -12pt
\caption{(color online): Absolute value of the decoherence factor
$|\nu_{\beta}(t)|$ (left panels, \textbf{A},\textbf{C}) and
excitation spectrum $\tilde{\nu}_{\beta}^{\prime }(E)$ (right
panels, \textbf{B},\textbf{D}), calculated from Eq.~(\ref{Vpbet})
by numerically computing the Gaussian damping~(\ref{eq:gbeta}) and
the periodic sub-diagram~(\ref{Lambda2E}), for $\protect\beta\hw
=1,3$, $\rF=5$, and $\protect\alpha =0.1-1.05$. }
\label{vtvbetrf5}
\end{figure}

\section{Decoherence factor, excitation spectrum and Non markovianity measure}

With the arguments given in the previous section, the decoherence
factor takes the analytical approximation
\begin{align*}
| \nu _{\beta }(t)| =&e^{-2 \alpha \omega^{2}t^{2}
\sum_{m=1}^{\infty }(-1)^{m} m e^{-\beta\hw m/2} }\\
&\times\prod_{m=-\infty }^{\infty}\left| \frac{e^{2\tau _{m}\omega
}-1}{e^{2(it+\tau _{m})\omega }-1}\right| ^{\alpha },
\end{align*}
 and the excitation spectrum relative to the \emph{perturbed} equilibrium energy of the gas may be written
\begin{equation*}
\tilde{\nu}^{\prime}_{\beta }(E)=\int_{0}^{\infty }\frac{dt}{2\pi
\hbar }e^{\frac{it}{\hbar }(E+E_1^{\beta}+E_2^{\beta})}\nu
^{\prime}_{\beta }(t).
\end{equation*}
We performed numerical calculations of both $| \nu_{\beta }(t) |$
and $\tilde{\nu}^{\prime}_{\beta }(E)$ by selecting different
fermion numbers~($\rF=5-100$), coupling parameters~($\alpha
=0.1-1.05$), and thermal energies~($\beta\hw =0.001-10$). Then, we
evaluated the maxima/minima of $|\nu _{\beta }(t)|$ in a finite
time-window with $t=0-100 \delta_{\beta}^{-1}$ and plugged the
differences $|\nu _{\beta }(t_{{\rm max},n})|-|\nu _{\beta
}(t_{{\rm min},n})|$ into Eq.~\eqref{Ndef}  to estimate the
non-Markovianity $\mathcal{N}$ of the two-level impurity in the
gas.
\begin{figure}[h]
\scalebox{0.99}{\includegraphics{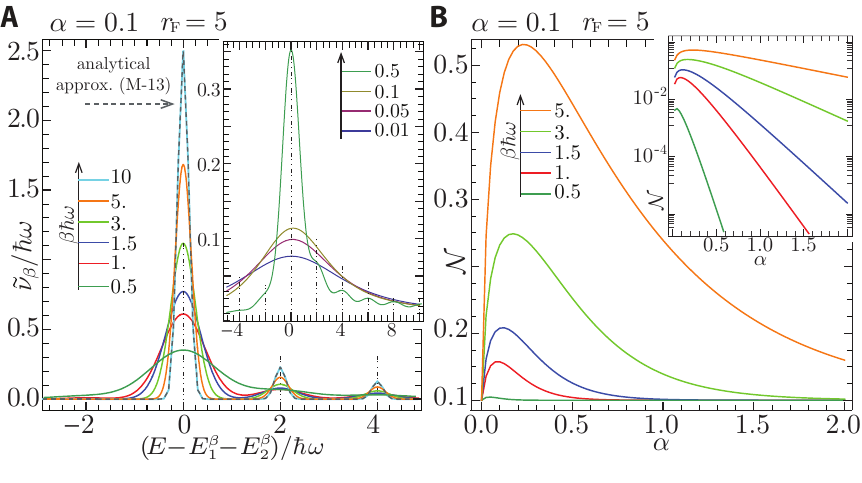}} \vskip -12pt
\caption{(color online): \textbf{(A)}~Absorption spectrum
$\nu^{\p}_{\beta}(E)$, calculated numerically from
Eq.~(\ref{Vpbet}) with $\beta\hw=0.1-\infty$, $\rF=5$, and $\alpha
=0.1$, and analytical approximation $\nu_{\infty }^{\p}(E)$
obtained from Eqs.~(\ref{Lambda2GA}) and~(\ref{LambXX}) with $\hw
\beta = 10$ and $\omega\tau_0 = 0.016$;
\textbf{(B)}~Non-Markovianity measure as a function of the
critical parameter $\alpha$ for various temperatures.}
\label{nonmarkB} \vskip -8pt
\end{figure}

In Fig.~3 and~4\textbf{A}, we have presented an application to a
Fermi gas of $402$ particles~($\rF=100$) where the Fermi-edge
behavior, superimposed on a Gaussian damping trend, appears as a
sequence of spikes in $|\nu _{\beta }(t)|=|\nu^{\p}_{\beta }(t)|$
and an asymmetric peak structure in $\tilde{\nu}^{\p}_{\beta
}(E)$. Such features becoming more and more marked with decreasing
temperature, which reduces the effect of the Gaussian damping
$\delta_{\beta }$~(see Fig.~2\textbf{A}), correspond to a sharp
peak of $\mathcal{N}$ at $\alpha<0.2$~(Fig.~4\textbf{B}). Similar
considerations hold for environments containing low fermion
numbers~(i.e, for $\rF=5$ in Fig.~\ref{vtvbetrf5} and
Fig.~\ref{nonmarkB}\textbf{A}) where shake-up effects are even
more visible because of the decreasing of $\delta_{\beta }$ with
decreasing $\eF$, leading to a sharp peaks in $\mathcal{N}$ at
$\alpha<0.5$~(Fig.~~\ref{nonmarkB}\textbf{B}). We finally notice
that the analytical the approximation given by
Eqs.~(\ref{Lambda2GA}) and~(\ref{LambXX}) works extremely well
with environments containing both
low~(Fig.~\ref{nonmark}\textbf{A}) and
large~(Fig.~\ref{nonmarkB}\textbf{A}) fermion numbers.

\end{document}